\documentclass[showpacs,twocolumn,aps]{revtex4}%
\usepackage{amsmath}
\usepackage{graphicx}
\usepackage{amsfonts}
\usepackage{amssymb}%
\begin{document}
\title{Efficient generation of universal two-dimensional cluster states with hybrid systems}
\author{Qing Lin}
\email{qlin@mail.ustc.edu.cn}
\affiliation{College of Information Science and Engineering, Huaqiao University (Xiamen),
Xiamen 361021, China}
\author{Bing He}
\email{heb@ucalgary.ca}
\affiliation{Institute for Quantum Information Science, University of Calgary, Alberta T2N
1N4, Canada}

\pacs{03.67.Lx, 42.50.Ex}

\begin{abstract}
We present a scheme to generate two-dimensional cluster state efficiently.
The number of the basic gate---entangler---for the operation is in the order of
the entanglement bonds of a cluster state, and could be reduced greatly if one
uses them repeatedly. The scheme is deterministic and uses few ancilla
resources and no quantum memory. It is suitable for large-scale quantum
computation and feasible with the current experimental technology.

\end{abstract}
\maketitle

\section{Introduction}

Measurement-based quantum computation (MBQC) or one-way quantum computation,
which was firstly introduced by Briegel and Raussendorf \cite{cluster}, has
been a hot topic in quantum information science recently. Different from the
traditional circuit-based quantum computation implemented by single-qubit and
multi-qubit gates, the necessary operation in MBQC is only projection
measurement on single qubits. However, the resources for MBQC should be the
entangled states of large numbers of qubits, which are conventionally called
cluster state or graph state. The efficient generation of such entangled
states is the main obstacle to the realization of MBQC. Many proposals have
been put forward for creating cluster states in various physics systems. They
include the discrete \cite{Nielsen1, Browne, dc-c} and continuous variable
\cite{CV-c} optical systems, the solid state systems such as charge qubits
\cite{charge}, flux qubit \cite{flux}, quantum dot \cite{dot}, etc.

Here we focus on the optical approaches to creating cluster states. In 2004,
Nielsen proposed the method of adding photons one by one with controlled-Z
(CZ) gates in generating a cluster state \cite{Nielsen1}. This scheme only
uses linear optical elements, so it is probabilistic and the cost for creating
a cluster state of large number of qubits could be very high. Later, many
works were developed to generate cluster state more efficiently. One of them
is the Browne-Rudolph protocol \cite{Browne}. Two types of fusion gates are
introduced in their protocol. The type-I fusion gate is used to connect two
cluster state strings with a success probability 1/2. After the operation of
this gate, an undetected photon will be connected to the photon adjacent to a
detected photon. The success of the gate is heralded by the detection of one
photon, i.e. one photon must be sacrificed. On the other hand, by a type-II
fusion gate, two photons are detected to create an L-shape cluster. At least
three photons must be consumed (one photon for the $\sigma_{x}$ operation, and
two photons for the Type-II fusion gate) in a complete operation. The
Browne-Rudolph protocol only applies linear optics, but it is also
nondeterministic and requires large quantity of sources (single photons), so
it is not appropriate for large-scale quantum computation.

More recently, another approach to generating cluster states with weak
nonlinearities were developed by S. G. R. Louis \textit{et al.} \cite{Spiller}%
. With the $\hat{X}$-quadrature measurements, its cluster state generation
could be deterministic, but the necessary amplitude of the measured coherent
beams should be $\alpha\theta^{2}\gg1$ ($|\alpha|$ is the amplitude of the
input coherent state, and $\theta$ the cross phase shift), so giant
nonlinearities should be demanded. If one chooses the $\hat{P}$-quadrature
measurements, the scaling will reduce to $\alpha\theta\gg1$, but it is
non-deterministic with a success probability 1/2. Moreover, their schemes
require a minus cross phase shift $-\theta$, which is impractical \cite{Kok}.

Besides the theoretical proposals, MBQC were experimentally demonstrated with
optical systems \cite{walther, LOE}, but it is impossible to follow these
proof-of-principle experiments to perform the realistic MBQC because quantum
memories are also necessary for storing the photonic cluster states. One could
generate cluster state with probabilistic gate, e.g., probabilistic controlled
phase flip gate \cite{CPF}, but it takes time to succeed in generating a whole
cluster state by the repeated gate operations. The already generated part of
the cluster state should be stored in quantum memories. If the efficiency of
generation is not high enough, a large number of photonic qubits in the
cluster state have to be kept in quantum memory for long time. Unfortunately,
efficient and faithful quantum memories for photonic qubits are still under
development thus far. Therefore, it is interesting to study how to quickly
create photonic cluster states without quantum memory.

In this paper, we propose a scheme to generate 2D cluster state with hybrid
systems involving discrete qubits and continuous variable states. This is a
deterministic approach to generating photonic cluster state of large size.
With the high efficiency of the scheme, only temporary storage such as delay
lines would be necessary for the involved operations.

The rest of the work is organized as follows. First, we describe a hybrid
system called entangler as the tool for creating the links of a cluster state.
Then, in Sec. III and IV, we outline the procedures to generate a string
cluster state and two types of box cluster state, respectively. Next, we
present the main results about the generation of 2D cluster state in Sec. V.
Finally we conclude the work with some discussions.

\section{Basic tool---entangler}

\begin{figure}[ptb]
\includegraphics[width=6.5cm]{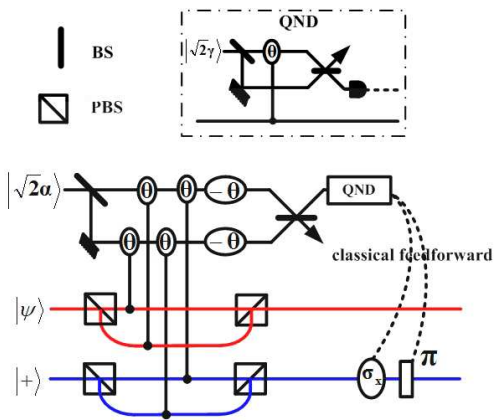}\caption{(Color online) Schematic
setup for entangler. Two qubus beams are coupled to the two single photons as
indicated. The XPM phases $\theta$ and two phase shifters $-\theta$ are
applied to the qubus beams. The QND module and the classical feedforward are
used to make this operation to be deterministic. After that, an entangled
state will be achieved.}%
\end{figure}

Before we present the scheme for generating cluster state, we describe the
tool of us to create the entanglement links in a cluster state. This gate is
called entangler briefly. It was first introduced by Pittman \textit{et al.,}
and then used to construct a CNOT gate \cite{Pittman}. Later, Nemoto
\textit{et al.} applied cross phase modulation (XPM) to make this gate
deterministic \cite{Nemoto}. Considering the impossibility of minus XPM phase
shift $-\theta$ in the scheme of \cite{Nemoto}, we used the technique of
double XPM to make such operation feasible with XPM \cite{Lin1}. The schematic
setup of our entangler is shown in Fig.1. The effect of the entangler is to
map the product of the states $\left\vert \psi\right\rangle =\alpha\left\vert
0\right\rangle +\beta\left\vert 1\right\rangle $ and $\left\vert
+\right\rangle =\frac{1}{\sqrt{2}}\left(  \left\vert 0\right\rangle
+\left\vert 1\right\rangle \right)  $ as follows:
\begin{equation}
\left\vert \psi\right\rangle \left\vert +\right\rangle \overset{E}%
{\rightarrow}\alpha\left\vert 0\right\rangle \left\vert 0\right\rangle
+\beta\left\vert 1\right\rangle \left\vert 1\right\rangle ,
\end{equation}
where $E$ denotes the entangler operation. The two input states will be thus
entangled, and the entangled output state inherits the coefficients of the
input $\left\vert \psi\right\rangle $. As shown in \cite{Lin1}, the error
probability in the operation is
\begin{equation}
P_{E}\sim exp\{-2(1-e^{-\frac{1}{2}\eta\gamma^{2}\theta^{2}})\alpha^{2}%
\sin^{2}\theta\},
\end{equation}
where $\alpha,\gamma$ are the amplitude of the coherent states used in the
operation and quantum non-demolition (QND) module for number-resolving
detection, respectively. Here $\theta$ is the XPM phase shift, while $\eta$ is
the efficiency of the detector. Even if $\theta\ll1$ for a weak cross-Kerr
nonlinearity, the operation would be deterministic given $\left\vert
\alpha\right\vert \sin\theta\gg1$ and $\eta\gamma^{2}\theta^{2}\gg1$. It
improves on the efficiency and feasibility of the entangler proposed in Refs.
\cite{Spiller}. Moreover, the QND module could be implemented with photon
number non-resolving detectors, such as APDs \cite{Lin1}.

Using two entanglers and one ancilla photon, a deterministic CNOT gate or CZ
gate can be realized \cite{Pittman, Nemoto}. Alternatively, one can also use a
pair of so-called C-path and Merging gate together with a recyclable ancilla
photon to realize the gates \cite{Lin1, Lin2}.

\section{Generation of string cluster state by Entangler}

As we know, a cluster string can be generated using CZ gates one by one
\cite{cluster, Nielsen1, Nielsen2}. However, this way may not be efficient. In
fact, using only one entangler is enough for generating a cluster string. We
begin with two initial states $\left\vert +\right\rangle ,\left\vert
+\right\rangle $. If a CZ gate is implied on these two states, we will get the
state $\frac{1}{\sqrt{2}}\left(  \left\vert 0\right\rangle \left\vert
+\right\rangle +\left\vert 1\right\rangle \left\vert -\right\rangle \right)
$, which is a two-qubit cluster state. Since the target state is a special
single photon state $\left\vert +\right\rangle $, just one entangler could let
us to achieve the two-qubit cluster state. The processes can be described as
follows,
\begin{align}
&  \frac{1}{\sqrt{2}}\left\vert +\right\rangle \left\vert +\right\rangle
\overset{E}{\rightarrow}\frac{1}{\sqrt{2}}\left(  \left\vert 0\right\rangle
\left\vert 0\right\rangle +\left\vert 1\right\rangle \left\vert 1\right\rangle
\right)  \nonumber\\
&  \overset{H_{2}}{\rightarrow}\frac{1}{\sqrt{2}}\left(  \left\vert
0\right\rangle \left\vert +\right\rangle +\left\vert 1\right\rangle \left\vert
-\right\rangle \right)  ,
\end{align}
where $H_{2}$ denotes the Hadamard gate performed only on the second photon.
Using only one entangler and one Hadamard gate, we can obtain a two-qubit
cluster state.

This method can be also used to add one photon to an already generated cluster
state, as shown in part (3) of Fig. 2. In general, an already existing cluster
state can be described in the unnormalized form $\left\vert \Phi
_{1}\right\rangle \left\vert 0\right\rangle _{p}+\left\vert \Phi
_{2}\right\rangle \left\vert 1\right\rangle _{p}$, where the other photons
except the $p$-th photon could be in an arbitrary state $\left\vert
\Phi_{1(2)}\right\rangle $. Now the process of adding one photon $q$ in the
state $\left\vert +\right\rangle $ to the already prepared cluster state is as
follows,
\begin{align}
&  ~~~~\left(  \left\vert \Phi_{1}\right\rangle \left\vert 0\right\rangle
_{p}+\left\vert \Phi_{2}\right\rangle \left\vert 1\right\rangle _{p}\right)
\left\vert +\right\rangle _{q}\nonumber\\
&  \overset{E_{pq}}{\rightarrow}\left\vert \Phi_{1}\right\rangle \left\vert
0\right\rangle _{p}\left\vert 0\right\rangle _{q}+\left\vert \Phi
_{2}\right\rangle \left\vert 1\right\rangle _{p}\left\vert 1\right\rangle
_{q}\nonumber\\
&  \overset{H_{q}}{\rightarrow}\left\vert \Phi_{1}\right\rangle \left\vert
0\right\rangle _{p}\left\vert +\right\rangle _{q}+\left\vert \Phi
_{2}\right\rangle \left\vert 1\right\rangle _{p}\left\vert -\right\rangle
_{q}.
\end{align}
The final result is the target cluster state.

Using this technique, one can easily generate any cluster state strings as
shown in part (1) of Fig. 2, and the star cluster state shown in part (2) of
Fig. 2. In addition, one can only use entanglers to generate an alveolate
graph shape deterministically ( the projector of the PBS in Ref. \cite{graph}
is actually an entangler) and a cluster state string simultaneously like the
scheme in \cite{string}. Another advantage is that no ancilla single phton is
necessary in the operation of entanglers. By the way, it should be noted that,
if one wants to connect two photons in two different already created cluster
states, one CZ gate, or two entanglers plus one ancilla single photon
equivalently, will be needed.

\begin{figure}[ptb]
\includegraphics[width=6cm]{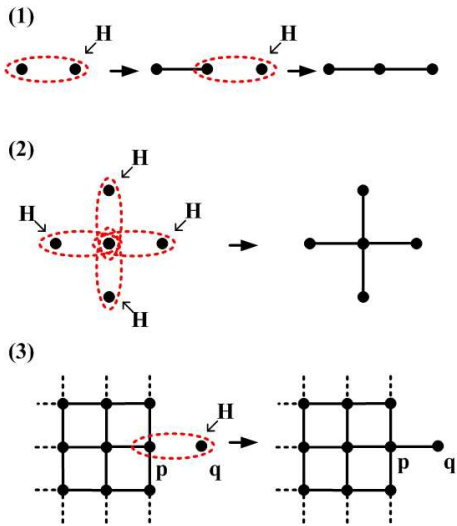}\caption{(Color online) The generation
of a cluster state string with entanglers. (1) Using entanglers and Hadamard
operations to generate a cluster state string. (2) Using four entanglers and
Hadamard operations to create a star cluster state. (3) using one entangler
and Hadamard operation to add one photon to the already created cluster
state.}%
\end{figure}

\section{Generation of two-dimensional box cluster state by Entangler}

A cluster state for practical MBQC should be of two-dimension ($2D$)
structure. Of course, one could use CZ gates to connect cluster state strings
to obtain a general $2D$ cluster state. However such practice could be still
not efficient enough. In what follows, we will first show how to generate a
box cluster state using a few entanglers, and then use the box cluster states
as the basic elements to construct a general $2D$ cluster state in an
efficient way.

\subsection{Type-I box}

The first scheme to generate a box cluster state is shown in part (1) of Fig.
3. At the beginning, we use two entanglers to generate a cluster state string
of three photons, which is described by the following state,
\begin{equation}
\frac{1}{2}\left(  \left\vert 0\right\rangle \left\vert +\right\rangle
\left\vert 0\right\rangle +\left\vert 0\right\rangle \left\vert -\right\rangle
\left\vert 1\right\rangle +\left\vert 1\right\rangle \left\vert -\right\rangle
\left\vert 0\right\rangle +\left\vert 1\right\rangle \left\vert +\right\rangle
\left\vert 1\right\rangle \right)  _{123}.
\end{equation}
Then, performing a Hadamard operation on the second photon, we will get%
\begin{equation}
\frac{1}{2}\left(  \left\vert 0\right\rangle \left\vert 0\right\rangle
\left\vert 0\right\rangle +\left\vert 0\right\rangle \left\vert 1\right\rangle
\left\vert 1\right\rangle +\left\vert 1\right\rangle \left\vert 1\right\rangle
\left\vert 0\right\rangle +\left\vert 1\right\rangle \left\vert 0\right\rangle
\left\vert 1\right\rangle \right)  _{123}.
\end{equation}
Next, applying an entangler operation on photon 2 and the photon 4 (initially
in the state $\left\vert +\right\rangle $) yields the state
\begin{align}
&  \frac{1}{2}\left(  \left\vert 0\right\rangle \left\vert 0\right\rangle
\left\vert 0\right\rangle \left\vert 0\right\rangle +\left\vert 0\right\rangle
\left\vert 1\right\rangle \left\vert 1\right\rangle \left\vert 1\right\rangle
\right. \nonumber\\
&  \left.  +\left\vert 1\right\rangle \left\vert 1\right\rangle \left\vert
0\right\rangle \left\vert 1\right\rangle +\left\vert 1\right\rangle \left\vert
0\right\rangle \left\vert 1\right\rangle \left\vert 0\right\rangle \right)
_{1234}.
\end{align}
Finally, a Hadamard operation is performed on the second and fourth photon,
respectively, and we will obtain the following state:%
\begin{align}
\frac{1}{2}  &  \left(  \left\vert 0\right\rangle \left\vert +\right\rangle
\left\vert 0\right\rangle \left\vert +\right\rangle +\left\vert 0\right\rangle
\left\vert -\right\rangle \left\vert 1\right\rangle \left\vert -\right\rangle
\right. \nonumber\\
&  \left.  +\left\vert 1\right\rangle \left\vert -\right\rangle \left\vert
0\right\rangle \left\vert -\right\rangle +\left\vert 1\right\rangle \left\vert
+\right\rangle \left\vert 1\right\rangle \left\vert +\right\rangle \right)
_{1234},
\end{align}
which is a box cluster state \cite{walther}. In this process, we
generate two bonds ($4\rightarrow1,4\rightarrow3$) simply by one
entangler operation and three Hadamard operations. The reason why
the operation could be thus simplified with entangler operation is
that the box cluster state has a perfect symmetry. Seen from photon
1 or 3, photon 2 and 4 are symmetric, so the states of them are
equivalent and one entangler operation will be enough for connecting
both bonds. Totally, 3 entangler operations, not 4 CZ gates, will be
necessary to generate a box cluster state. No ancilla photon is
needed for the entangler operations here.

Generalizing the scheme to adding a box cluster state to an already generated
cluster state is straightforward. The schematic setups are shown in part (2)
of Fig. 3. Generally, an already created cluster state is in the unnormalized
form $\left\vert \Phi_{1}\right\rangle \left\vert 0\right\rangle
_{p}+\left\vert \Phi_{2}\right\rangle \left\vert 1\right\rangle _{p}$. First
we apply the procedure in Eq. (3) to add one photon ($\left\vert
+\right\rangle _{q}$) to an already created cluster state to get the state
$\left\vert \Phi_{1}\right\rangle \left\vert 0\right\rangle _{p}\left\vert
+\right\rangle _{q}+\left\vert \Phi_{2}\right\rangle \left\vert 1\right\rangle
_{p}\left\vert -\right\rangle _{q}$. Secondly, continuing to add one more
photon ($\left\vert +\right\rangle _{r}$) to obtain the state
\begin{align}
&  \frac{1}{\sqrt{2}}\left[  \left\vert \Phi_{1}\right\rangle \left\vert
0\right\rangle _{p}\left(  \left\vert 0\right\rangle _{q}\left\vert
+\right\rangle _{r}+\left\vert 1\right\rangle _{q}\left\vert -\right\rangle
_{r}\right)  \right. \nonumber\\
&  \left.  +\left\vert \Phi_{2}\right\rangle \left\vert 1\right\rangle
_{p}\left(  \left\vert 0\right\rangle _{q}\left\vert +\right\rangle
_{r}-\left\vert 1\right\rangle _{q}\left\vert -\right\rangle _{r}\right)
\right] \nonumber\\
&  =\frac{1}{\sqrt{2}}\left[  \left\vert \Phi_{1}\right\rangle \left\vert
0\right\rangle _{p}\left(  \left\vert +\right\rangle _{q}\left\vert
0\right\rangle _{r}+\left\vert -\right\rangle _{q}\left\vert 1\right\rangle
_{r}\right)  \right. \nonumber\\
&  \left.  +\left\vert \Phi_{2}\right\rangle \left\vert 1\right\rangle
_{p}\left(  \left\vert -\right\rangle _{q}\left\vert 0\right\rangle
_{r}+\left\vert +\right\rangle _{q}\left\vert 1\right\rangle _{r}\right)
\right]  .
\end{align}
Finally, by a similar process shown from Eq. (5) to Eq. (8), we can achieve
the following state
\begin{align}
\frac{1}{\sqrt{2}}  &  \left\vert \Phi_{1}\right\rangle \left(  \left\vert
0\right\rangle \left\vert +\right\rangle \left\vert 0\right\rangle \left\vert
+\right\rangle +\left\vert 0\right\rangle \left\vert -\right\rangle \left\vert
1\right\rangle \left\vert -\right\rangle \right)  _{pqrs}\nonumber\\
&  +\frac{1}{\sqrt{2}}\left\vert \Phi_{2}\right\rangle \left(  \left\vert
1\right\rangle \left\vert -\right\rangle \left\vert 0\right\rangle \left\vert
-\right\rangle +\left\vert 1\right\rangle \left\vert +\right\rangle \left\vert
1\right\rangle \left\vert +\right\rangle \right)  _{pqrs},
\end{align}
which is the desired cluster state. With only 3 entanglers, we can add a box
structure to an already created cluster state. Here only one photon in the
added box belongs to the already created cluster state, i.e., the added box
must include three photons which are not in the already generated cluster
states. We call this type of box cluster state as Type-I box.

\begin{figure}[ptb]
\includegraphics[width=6cm]{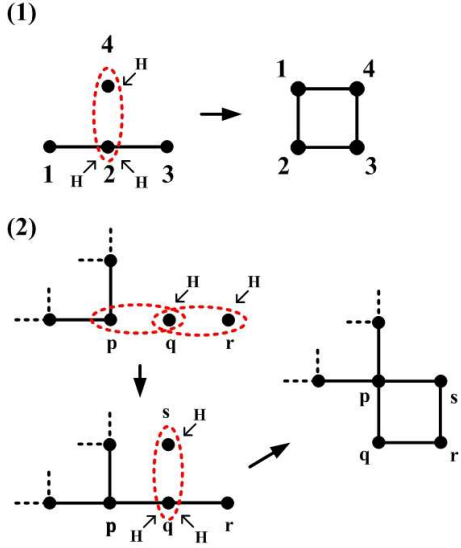}\caption{(Color online) Schematic setup
for generating Type-I box. (1) One entangler plus three Hadamard operations
are performed on photon 2 and 4 in order to connect two links, 1-4, 3-4. Then
a box cluster state could be generated by only three entangler operation
without ancilla photon. (3) The generalization of adding a box structure to an
already generated cluster state.}%
\end{figure}

\subsection{Type-II box}

Since the type-I box can only be used to add three photons to an already
generated cluster state, its application in generating a $2D$ cluster state is
limited. Here we introduce another type of box cluster state called Type-II
box (Fig.4). This box could be used to connect two photons which belong to an
already created cluster states (or two different cluster states) and two
independent photons. Suppose the already generated cluster state is initially
prepared as (unnormalized)
\begin{align}
&  \left\vert \Psi_{1}\right\rangle \left\vert 0\right\rangle _{p}\left\vert
0\right\rangle _{s}+\left\vert \Psi_{2}\right\rangle \left\vert 0\right\rangle
_{p}\left\vert 1\right\rangle _{s}\nonumber\\
&  +\left\vert \Psi_{3}\right\rangle \left\vert 1\right\rangle _{p}\left\vert
0\right\rangle _{s}+\left\vert \Psi_{4}\right\rangle \left\vert 1\right\rangle
_{p}\left\vert 1\right\rangle _{s}.
\end{align}
At first, we use two entanglers and some Hadamard operations to add two
photons respectively in the states $\left\vert +\right\rangle _{q},\left\vert
+\right\rangle _{r}$ to the above cluster state to get the following state%
\begin{align}
&  \frac{1}{\sqrt{2}}\left[  \left\vert \Psi_{1}\right\rangle \left\vert
0\right\rangle _{p}\left\vert 0\right\rangle _{s}\left(  \left\vert
+\right\rangle _{q}\left\vert 0\right\rangle _{r}+\left\vert -\right\rangle
_{q}\left\vert 1\right\rangle _{r}\right)  \right. \nonumber\\
&  +\left\vert \Psi_{2}\right\rangle \left\vert 0\right\rangle _{p}\left\vert
1\right\rangle _{s}\left(  \left\vert +\right\rangle _{q}\left\vert
0\right\rangle _{r}+\left\vert -\right\rangle _{q}\left\vert 1\right\rangle
_{r}\right) \nonumber\\
&  +\left\vert \Psi_{3}\right\rangle \left\vert 1\right\rangle _{p}\left\vert
0\right\rangle _{s}\left(  \left\vert -\right\rangle _{q}\left\vert
0\right\rangle _{r}+\left\vert +\right\rangle _{q}\left\vert 1\right\rangle
_{r}\right) \nonumber\\
&  \left.  +\left\vert \Psi_{4}\right\rangle \left\vert 1\right\rangle
_{p}\left\vert 1\right\rangle _{s}\left(  \left\vert -\right\rangle
_{q}\left\vert 0\right\rangle _{r}+\left\vert +\right\rangle _{q}\left\vert
1\right\rangle _{r}\right)  \right]  .
\end{align}
Next, after a Hadamard operation is performed on photon $q$, we perform a CZ
operation on photon $q$ and $s$, respectively. Finally, a Hadamard operation
on photon $q$ will achieve the state%
\begin{align}
&  \frac{1}{\sqrt{2}}\left[  \left\vert \Psi_{1}\right\rangle \left\vert
0\right\rangle _{p}\left\vert 0\right\rangle _{s}\left(  \left\vert
+\right\rangle _{q}\left\vert 0\right\rangle _{r}+\left\vert -\right\rangle
_{q}\left\vert 1\right\rangle _{r}\right)  \right. \nonumber\\
&  +\left\vert \Psi_{2}\right\rangle \left\vert 0\right\rangle _{p}\left\vert
1\right\rangle _{s}\left(  \left\vert +\right\rangle _{q}\left\vert
0\right\rangle _{r}-\left\vert -\right\rangle _{q}\left\vert 1\right\rangle
_{r}\right) \nonumber\\
&  +\left\vert \Psi_{3}\right\rangle \left\vert 1\right\rangle _{p}\left\vert
0\right\rangle _{s}\left(  \left\vert -\right\rangle _{q}\left\vert
0\right\rangle _{r}+\left\vert +\right\rangle _{q}\left\vert 1\right\rangle
_{r}\right) \nonumber\\
&  \left.  +\left\vert \Psi_{4}\right\rangle \left\vert 1\right\rangle
_{p}\left\vert 1\right\rangle _{s}\left(  -\left\vert -\right\rangle
_{q}\left\vert 0\right\rangle _{r}+\left\vert +\right\rangle _{q}\left\vert
1\right\rangle _{r}\right)  \right]  .
\end{align}
which is the target cluster state with the box structure for the photons
$p,q,r,s$. Two entangler operations and one CZ gate are necessary for
generating this Type-II box. Since a CZ gate could be implemented by two
entanglers, four entanglers will be enough to create this type of box. On
average, one entanglement bond needs one entangler operation by this method.

\begin{figure}[ptb]
\includegraphics[width=6cm]{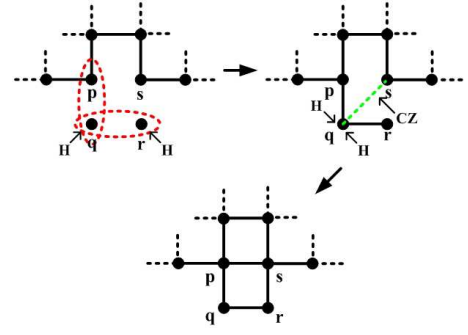}\caption{(Color online) Schematic setup
for generating Type-II box. Two entanglers are used to add two single photons
to the already generated cluster state. Then, one CZ gate is performed on
photon q and s for creating two links, s-r, s-p. Totally, four entanglers with
one ancilla photon are required to add a Type-II box between two already
created cluster states.}%
\end{figure}

\section{Creating a general 2-D cluster state with Entanglers and CZ gates}

In a classical computer, a simple computation task could involve thousands of
bits. Though numerous experiments in MBQC have shown the power of quantum
computation, all of them are proof of principle in nature \cite{walther, LOE}.
The quantities of qubits in these experiments are limited, and only simple
operations could be demonstrated. Highly efficient schemes of generating
cluster states must be developed before the large-scale computation in MBQC
could possibly materialize. As the main topic of the paper, we will show in
the following how to generate an arbitrary $2D$ cluster state using the above
discussed string, box cluster states as the basic elements.

We illustrate the procedure with the example of a $5\times5$ cluster state,
which is shown in Fig. 5. Six steps will complete the generation of this
cluster state:

(1) using 8 entangler operations, a cluster state string of 9 qubits is generated;

(2) applying the procedure of creating Type-I box for four times, a cluster
state of four boxes will be achieved;

(3) adding two cluster state strings of 4 qubits to the second box, and then
two Type-I box will be added to the four-box cluster state with two entanglers;

(4) continuing to add two type-II box to the six-box cluster state with two
cluster state strings and two CZ gates;

(5) adding two independent photons to the eight-box cluster state
with two entanglers;

(6) finally, six CZ gates will connect the links to the target $5\times5$
cluster state.

\begin{figure}[ptb]
\includegraphics[width=6cm]{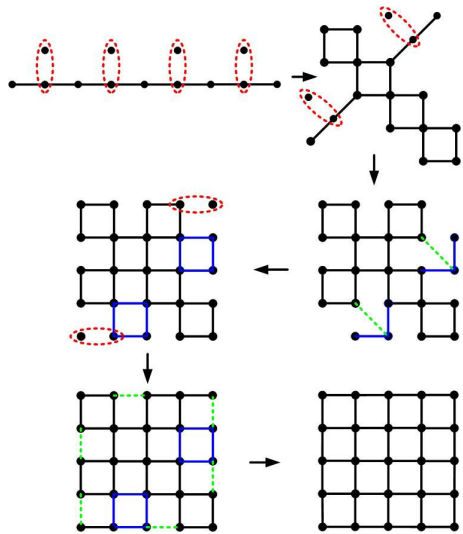}\caption{(Color online) The generation of a
$5\times5$ cluster state from string and two types of box
structures. A 9-photon cluster string and four entanglers are used
to create four Type-I box cluster states, and then two Type-I boxes
are added by six entanglers. Next, two Type-II boxes generated by
four entanglers and two CZ gates will be used. After that, with two
entanglers, two independent photons will be added to the eight-box
cluster state. At the last step, six CZ gates are applied to
complete the generation. There are totally 40 necessary entanglers.
The number of entanglers is equal to that
of the bonds of the cluster state.}%
\end{figure}

Here we neglect the use of Hadamard operations for a simpler
illustration. Now, we calculate the resources required in this
scheme. Besides some single-qubit operations, 24 entanglers are
required in the generation of string, two types of box structures; 8
CZ gates are required in the generation of Type-II box and in the
final step. Considering the fact that one CZ gate could be realized
by two entanglers, totally 40 entanglers should be used in this
scheme. The number of entanglers is exactly equal to the bonds of
the $5\times5$ cluster state.

It is straightforward to generalize this method to create an $n\times n$
cluster state in the approach. If $n$ is odd, $n^{2}-1$ entanglers and
$\left(  n-1\right)  ^{2}/2$ CZ gates, or totally $2n\left(  n-1\right)  $
entanglers will be required in the generation of a $n\times n$ cluster state.
If $n$ is even, $n^{2}-1$ entanglers and $n\left(  n-2\right)  /2$ CZ gates,
or totally $2n\left(  n-1\right)  -1$ entanglers will be necessary to generate
the cluster state. Evidently, the number of the entanglers is less than or
equal to the number of the bonds. In other words, we could generate a
universal $2D$ cluster state with one entangler operation per bond, so the
scheme is highly efficient.

\section{Discussion and conclusion}

In this paper, we propose a scheme to generate a general $2D$ cluster state
with entanglers plus a few CZ gates. In their realizations with linear optics
and weak nonlinearity \cite{Lin1}, entangler operations need no ancilla single
photon and a CZ gate operation uses one ancilla photon which could be recycled
if we use QND module in detection. It is also shown in \cite{Lin1} that QND
module could be realized with common photon number non-resolving detectors
such as APDs. Therefore, only one ancilla single photon is necessary for
creating a general $2D$ cluster state in principle, even if the cluster state
involves large number of qubits.

The number of the basic gate---entangler---required by the scheme is
in the order of the cluster state bond number, and the resources
will be greatly reduced if the entanglers are used repeatedly in
operation. Actually, this scheme works with the fixed circuits
(entanglers or CZ gates) with which the different links can be
generated by the same setup if the process is in time order. Like
the operations in a classical computer, only the simultaneous
operations require different circuit resources, so the number of the
necessary entanglers could be much smaller than the number of bonds
in a cluster state. We indicate the time order with the arrows in
figures, where the different steps could be done by the same
entangler, CZ gate, etc. Moreover, the system enjoys the advantage
of deterministic operation of entanglers, and it reduces the
generation time for clusters greatly. So the scheme is more suitable
for large-scale quantum computation than the previously proposed
ones.

As mentioned in the introduction, quantum memory is required in the
schemes with probabilistic gates, which repeats the operation until
success. The already generated parts should be stored in memory.
However, the realization of high-quality quantum memory\ is still
technically challenging. In our scheme, the entangler operation
based on XPM is deterministic and very fast (the operation time is
in the order of that for the signals going through the nonlinear
medium). So the storage time for the already generated parts need
not to be long, and one could use some temporary storage, such as
delay lines, for the already created parts. In this sense, the
scheme could be feasible with the current experimental technology.

The scheme improves on the previous ones by replacing the "one by one" fashion
of generating the entanglement links with the "string by string" and "box by
box", which increase the efficiency greatly. In particular, this approach to
MBQC is deterministic, and uses less resources and no quantum memory. It could
be a promising candidate for large-scale quantum computation.

\begin{acknowledgments}
The authors would like to thank Dr. Peter van Loock and Ru-Bing Yang
for helpful suggestions. B.H. acknowledges the support by $i$CORE.
\end{acknowledgments}

\end{document}